\documentclass{PoS}

\usepackage{amsmath}
\usepackage{amssymb}
\usepackage{url}
\usepackage{graphics} 
\usepackage{bm} 

\usepackage{layouts}

\usepackage[above]{placeins}
\usepackage{framed}
\usepackage{subfig}

\newcommand{\braket}[2]{\langle#1|#2\rangle}

\newcommand{\mmat}[4]{\ensuremath{\begin{pmatrix} {#1} & {#2} \\ {#3} &
    {#4} \end{pmatrix}}}

\title{Fast Evaluation of Multi-Hadron Correlation Functions}

\ShortTitle{Fast Evaluation of Multi-Hadron Correlation Functions}

\author{\speaker{Pranjal Vachaspati}\\
        Massachusetts Institute of Technology\thanks{Currently at the
          University of Illinois at Urbana-Champaign}\\
        E-mail: \email{vachasp2@illinois.edu}}

\author{William Detmold\\
        Massachusetts Institute of Technology\\
        E-mail: \email{wdetmold@mit.edu}}

      \abstract{Calculating the values of nuclear correlation
        functions is computationally intensive due to the fact that
        the number of terms in a nuclear wave function scales
        exponentially with atomic number. To speed up this
        computation, we represent a correlation function as a sum of
        the determinants of many small matrices, and exploit
        similarities between the matrices to speed up the calculations
        of those determinants. }

\FullConference{The 32nd International Symposium on Lattice Field Theory,\\
		23-28 June, 2014\\
		Columbia University New York, NY}
\newcommand{\N}{\mathcal N}

\begin{document}

\section{Introduction}

We wish to find the correlation function between a source and a sink
nuclear wave function, of the form 
  \begin{equation}
    \braket{\N_1(t)}{\bar \N_2(0)}     
  \end{equation}
  where $\bar \N_h$ and $\N_h$ are nuclear creation and annihilation
  operators built from quark and gluon fields described by quantum
  numbers $h$. As described in \cite{detmold2013nuclear}, $\bar \N_h$
  can be written as
  \begin{equation}
    \bar \N_h = \sum_{\{a\}} w_h^{a_1\ldots a_{n_q}}\bar q(a_1) \bar q(a_2) \ldots
    \bar q(a_{n_q})
  \end{equation}
  where $w_h$ is an antisymmetric tensor that is nonzero only when
  none of the indices are equal and the combination of the quarks
  is appropriate for the nuclear quantum number.

  The correlation function requires a sum over pairs of terms in the
  source and sink wavefunctions, and each pair requires a sum over
  orderings of quark creation and annihilation operators in each term.
  It can be expressed in terms of quark propagators $S(q(a_0), q(a_1);
  t)$ as: 
  \begin{equation}
      \braket{\N_1(t)}{\bar \N_2(0)} = \frac{1}{Z} \int d{U}
      e^{-S_{QCD}} \sum_{i \in src} \sum_{j \in snk} w_{i} w_{j}
      \sum_{{\alpha},{\beta}} S(\bar q({a_i}^{\alpha_1}), q({a_j}^{\beta_1}))
      \ldots S(\bar q({a_i}^{\alpha_{nq}}), q({a_j}^{\beta_{nq}}))
      \epsilon^{\alpha} \epsilon^{\beta}
  \end{equation}
  where $\alpha$ and $\beta$ are permutations of $1 \ldots nq$, and $a_n^m$ is
  the $m^{th}$ creation or annihilation operator in the $n^{th}$ term
  of the wavefunction.
  
  For a given source term $i$ and sink term $j$, the sum over $\alpha$ and
  $\beta$ is equivalent to taking the determinant of a matrix $G^{(ij)}$
    where \cite{detmold2013nuclear}
  \begin{equation}
    G_{mn}^{(ij)} = S(\bar q(a_i^m), q(a_j^n))
  \end{equation}

  Therefore, the expression for the correlator can be restated as:
  \begin{equation}
      \braket{\N_1(t)}{\bar \N_2(0)} = \frac{1}{Z} \int d{U}
      e^{-S_{QCD}} \sum_{i \in src} \sum_{j \in snk} w_{i} w_{j}
      \det({G^{(ij)}})
  \end{equation}  
  Here, computing the nuclear correlation function requires computing
  a determinant for every pair of terms in the sink and source wave
  functions, which typically grow rapidly in size with increasing
  atomic number $A$. Note that $G$ is block-diagonal because the
  correlation function between two quarks of different flavors is
  always zero, so we compute $\det(G)$ as the product of the
  determinants of smaller single-flavor matrices.
\section{Determinant Updates}
Typically, $\det(G)$ is calculated by taking the trace of the $LU$
decomposition of the matrix, which takes $O(n^3)$ time in the number
of quarks \cite{golub2012matrix}. However, this can be improved by
taking into account similarities between pairs of matrices.

If two source wavefunctions or two sink wavefunctions differ by only
one term, the corresponding $G$ matrices will differ by only a single
row or column. If source terms $i$ and $j$ differ only in the $r^{th}$
element, the matrices $G^{(ik)}$ and $G^{(jk)}$, which denote,
respectively, the correlator between the $i^{th}$ or $j^{th}$ source
term and the $k^{th}$ sink term, can be expressed as:
\begin{equation}
G^{(jk)} = G^{(ik)} + uv^T
\end{equation}
where $v$ is the difference between the rows of $G^{(jk)}$ and $G^{(ik)}$,
and $u$ is an element vector that determines which row has changed. 

\subsection{Matrix Determinant Lemma}
\label{sec:mdl}

An update of this form can be computed in $O(n^2)$ time, using the
identity \cite{ding2007eigenvalues}
\begin{equation}
\det (G + uv^{T}) = (1 + v^T G^{-1} u)\det(G).
\end{equation}

To prove this, we first see that:
\begin{equation}
  \det (I + uv^{T}) = (1 + v^T u)
\end{equation}
because of the equality
\begin{equation}
  {\mmat 1 0 {v^T} 1} {\mmat {1 + uv^T} u 0 1} {\mmat 1 0 {-v^T} 1} = 
  {\mmat I u 0 {1 + v^T u}}
\end{equation}
This implies
\begin{equation}
\det(G+ uv^T) = \det(G)\det(I + G^{-1}uv^T) = \det(G)\det(I + v^TG^{-1}u)
\end{equation}

$G^{-1}u$ is easily computed using the $LU$ factorization, since the
systems $Ux = b$ and $Ly = x$ can be solved in a total of $n^2-n$
multiply-adds and $2n$ divides.

Somewhat surprisingly, it is possible to calculate some rank-1 updates
in $O(n)$ time. If a parent matrix $G^{(ik)}$ has the $n^{th}$ row
changed in one way to get $G^{(jk)}$ and in a different way to get
$G^{(lk)}$, the determinant of $G^{(jk)}$ is 
\begin{equation}
\det (G^{(ik)} + uv_1^{T}) = (1 + v_1^T {G^{(ik)}}^{-1} u)\det(G^{(ik)}),
\end{equation}
and the determinant of $G^{lk}$ will be
\begin{equation}
\det (G^{(ik)} + uv_2^{T}) = (1 + v_2^T {G^{(ik)}}^{-1} u)\det(G^{(ik)}).
\end{equation}
The most computationally intensive calculation is finding ${G^{(ik)}}^{-1}
u$, but this is shared between the two determinants. So the
determinant of $G^{(lk)}$ can be computed with only a single vector
multiply. 

\section{Execution Order}

To be able to use the fast determinant method, we must find a set $S$
of matrices such that the every matrix $G$ differs from some matrix in
$S$ by at most one row or column. Then, after evaluating $\det(G_1)$
where $G_1 \in S$, which takes $O(n^3)$ time, all the matrices that
differ from $G_1$ by a row or column can have their determinants
calculated in $O(n^2)$ or $O(n)$ time.

Consider a graph in which each matrix is a vertex and an edge exists
between vertices if they differ by a single row or column. The subset
$S$ is represented by a dominating set of this graph. Ideally, we
would use the minimal dominating set, as this would require the fewest
uses of the $O(n^3)$ algorithm.

However, finding the minimal dominating set of a graph is known to be
$NP$-complete. Luckily, a simple greedy algorithm, where the vertex with
the highest degree is chosen and removed from the graph, along with
its neighbors, provides a good approximation
\cite{parekh1991analysis}.

Furthermore, simply generating the graph is difficult, as $O(N^4)$
matrix comparisons are needed. For this reason, only the dominating
set of the source wavefunction is considered. This has complexity of
only $O(N^2)$, which takes no more than a few minutes given a wave
function with order $10^4$ terms. The graph and dominating set remain
the same over different gauge configurations, so the time spent in
this phase can be amortized over many correlator calculations.

In practice, the vertices in the graph tend to have high degree, so
a large fraction of the matrices can have their determinants
calculated in linear or quadratic time.

\section{Results}
\begin{figure}\centering
  
  \input{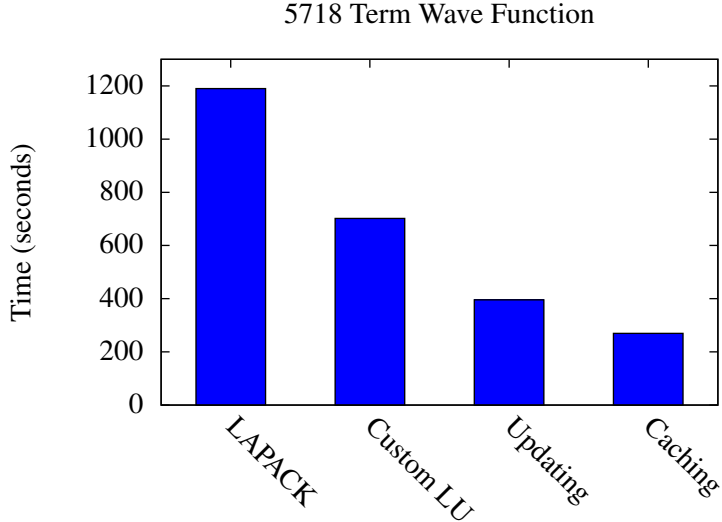}
  \caption{A comparison of algorithms shows that the fast calculation is
    approximately four times as fast as LAPACK. ``Updating'' uses the
    $O(n^2)$ update algorithm where it can, and ``caching'' additionally
    saves the solution of $G^{-1}u$ to compute a portion of the
    determinants in $O(n)$ time.}
  \label{fig:graph}
\end{figure}

This method was tested using five sample alpha particle wave
functions. One had all 12 quarks on the same site with one term; two
consisted of a triton and a proton, with 1944 terms each; and two had
two particles on each site, with 5718 terms each. All combinations of
wavefunctions at the sink and source were considered, so when both the
sink and source had 5718 terms, a total of $5718^2 = 32,695,524$
determinants were calculated per site.

The wave function with one term had a dominating set of size one; the
1944 term wave functions had dominating sets of size 167, 183, 241,
and 245; and the 5718 term wave functions had dominating sets of sizes
509, 568, 699, and 703. Thus, between 80 and 90 percent of the
determinants could be calculated more quickly than with the na\"ive
method. 

First, the LU factorization function in LAPACK was replaced with a
custom method that did not use pivoting, a technique that enhances
numerical stability for poorly conditioned matrices. LAPACK is
optimized for larger matrices; for the small ($6 \times 6$) matrices
used here, LAPACK has considerable overhead. The optimizations
discussed above were applied using this custom factorization function.

Overall, the determinant updating technique was between four and five
times faster than LAPACK when tested on combinations of the wave
functions above, as can be seen in Figure \ref{fig:graph}.

\section{Conclusion \& Further Improvements}

Improving two portions of the algorithm would have significant impacts
on efficiency. 

The first is the quality of the dominating set approximation, which
determines how many determinants must be computed in cubic time. A
simple greedy algorithm works reasonably well, but it may be
worthwhile to spend more time in this phase to get a better
result. Second, reducing overhead in the algorithm and in the caching
mechanism becomes increasingly important as the asymptotic complexity
of the determinant calculation decreases.

Larger nuclei would benefit more from this approach, since the
difference between the $O(n^3)$ approach and the faster methods would
be more significant. For sufficiently large nuclei, utilizing rank-2
and higher updates might be effective as well.

Additionally, this method can be parallelized up to the size of the
dominating set, as the graph can be partitioned into sets which each
contain one member of the dominating set and its neighbors. Each of
these partitions can be considered separately.

Utilizing the similarities between wavefunction terms can provide
significant speedups in the calculation of correlators. The matrix
determinant lemma provides an asymptotic speedup from $O(n^3)$ to
$O(n^2)$ for many of the terms, and careful caching provides a further
speedup to $O(n)$. This asymptotic speedup is reflected in real-world
correlator calculations.

\bibliographystyle{unsrt}
\bibliography{skeleton}

\end{document}